\RequirePackage{fix-cm}
\documentclass[twocolumn,epjc3]{svjour3}  
\smartqed  
\RequirePackage{graphicx}
\journalname{Eur. Phys. J. C}
\usepackage{cite}
\usepackage{amsmath}
\usepackage{amsfonts}
\usepackage{amssymb}
\usepackage{makeidx}
\usepackage{graphicx}
\usepackage{lmodern}
\usepackage{color}

\begin{document}
\title{\bf Gravitational radiation of a spherically symmetric source \\
in {\boldmath $f(R)$}-gravitation } 
%
\author{Pham Van  Ky\thanksref{e1}
        \and
Nguyen Thi Hong  Van\thanksref{e2}
        \and 
Nguyen Anh Ky\thanksref{e3}
}
\thankstext{e1}{e-mails: phamkyvatly AT gmail.com}
\thankstext{e2}{e-mail: nhvan AT iop.vast.vn} 
\thankstext{e3}{e-mail: anhky AT iop.vast.vn}
\institute
{\small 
\textit{Mathematical-, high energy- and astro-physics group, CTP,}\\
\small
\textit{Institute of physics},\\ 
\small
\textit{Vietnam academy of science and technology (VAST)},\\ 
\small
\textit{10 Dao Tan, Ba Dinh, Hanoi, Viet Nam.}
}
\date{Received: date / Accepted: date}
%
%

\maketitle
%
%
\begin{abstract}
It is shown that Birkhoff's theorem for the general theory of relativity is overcome in the $f(R)$-theory of gravitation.
That means, the $f(R)$-theory of gravitation, unlike Einstein's general theory of relativity, does not forbid  gravitational radiation from a spherically symmetric source (whether stationary  or non-stationary). As a consequence, in the $f(R)$-theory a spherically symmetric gravitational deformation (e.g., collapse/expansion or pulsation) could emit gravitational waves (of tensor- and scalar polarization modes), a phenomenon impossible in the general relativity. A test model is examined and it turns out that the gravitational radiation is strongest when the surface of the deforming object is in the vicinity of the (modified) event horizon, even suddenly flares up just outside the latter.  In this letter, within the $f(R)$-theory of gravitation, a gravitational wave equation and a formula for the gravitational emission power are  derived. These formulae, along with searching for signals, can be used for the experimental test of the $f(R)$-theory. In general, including the spherically symmetry case, gravitational radiation of both tensor- and scalar polarization modes are allowed, although under some circumstance the contribution of scalar modes is strongly suppressed.
\end{abstract}
\maketitle 

\section{Introduction}

As is well-known in the general theory of relativity (GR),  a solution of Einstein's equation in vacuum for a central field is always stationary (according to Birkhoff's theorem)  \cite{Birkhoff1,VojeJohansen:2005nd} as the corresponding metric can be converted into a time-independent form \cite{Weinberg1,Landau1}. 
Therefore, in the GR no gravitational waves can be emitted from any spherically varying  gravitational source including spherically collapsing (or expanding) and pulsating  ones\footnote{In practice there is no perfect spherical object but  deformation from the spherical shape may not always be strong and fast enough to emit detectable  gravitational waves.}. This statement, as shown below (see also 
\cite{KKV}), however, is no longer valid for an $f(R)$-theory of gravitation, or just an $f(R)$-gravitation or an $f(R)$-theory, for short. It has been shown \cite{ReddyEPJP2018} that Birkhoff's theorem still holds in the $f(R)$ theory only for specific cases, such as that with time-independent $R$, not applicable to the general $f(R)$ theory considered here.\\

As a result, a spherically symmetric deforming (pulsating/contracting/extending) object in a general 
$f(R)$-gravitation, unlike in the GR, may be able to emit gravitational waves detectable from  distance 
if the intensity and the speed of the deforming process are high enough. Thus, if an $f(R)$-gravitation \cite{DeFelice:2010aj,Sotiriou:2008rp,Capozziello:2011et}
replaces the Einstein's GR as a more precise theory of gravitation the number of possible sources of gravitational waves may increase, for example, a star in evolution (to a black hole or a neutron star, or a supernova, etc.), keeping even its spherical symmetry, could emit gravitational waves.
Here we will demonstrate theoretically how it can happen. In particular, a spherical gravitational collapse  (of a single object), not only gravitational collisions (between two black holes, or neutron stars, for example) \cite{LIGOScientific:2016aoc}, could be also a source of gravitational waves. This conclusion was reached thanks to the perturbation approach developed in \cite{KKV} for cosmic objects, in \cite{VanKy:2022itq} for the Universe as a whole and here for gravitational radiations. The perturbation approach in the $f(R)$-theory has been also discussed elsewhere  (see, e.g.,  \cite{Katsuragawa:2019uto, Gogoi:2020ypn}) 
but on the background of the perturbation of the metric as usual (see \eqref{metric} below), not of the theory itself.\\

In general, the GR admits tensor polarized gravitational radiation only, but a generic metric theory may admit more, until five or six, polarization modes of gravitational waves \cite{Eardley:1973br,Eardley:1973zuo}, in particular, the $f(R)$-theory admits gravitational waves from scalar modes in addition to those of the tensor modes  (see, for example, \cite{Capozziello:2008rq,Nishizawa:2009bf} and references therein). The presence of gravitational waves of non-tensor modes, including the scalar ones,  indicates non-GR phenomena. However, the observations of the Advanced LIGO and the Advanced Virgo have shown the signal of tensor modes dominates that of scalar modes. \cite{Isi:2017fbj, LIGOScientific:2018czr,Isi:2022mbx}. This is somewhat consistent with the theoretical results \cite{Inagaki:2023tjh}. Moreover, gravitational waves of scalar modes are more related to the massive gravitation rather than the massless one \cite{Inagaki:2023tjh}. In the spherically symmetry condition, as said above, no gravitational waves can be radiated within the GR, in contrast,  the gravitational radiation is still possible in the $f(R)$ theory under the spherically symmetry condition. We will show that they can be of tensor modes. Along with the would-be existence of non-tensor modes of gravitational waves, the would-be existence of gravitational waves of a tensor mode from a spherically symmetric (varying) source would also mean physics beyond the GR. However, as stated in \cite{Inagaki:2023tjh} the scalar modes of the gravitational waves are sometimes strongly suppressed 
by the mass correction.\\   

Let us outline the paper. In the next section we recall some key points from the derivation of the gravitational radiation in the GR and in Sect. 3 we consider this phenomenon for the $f(R)$ theory in a central field. In Sect. 4 we apply our approach to several test models and analyze the results with some illustrations.\\

The conventions used throughout this paper are the same as those used in our previous works \cite{KKV,VanKy:2022itq}.

%
\section{Gravitational radiation in the general relativity in brief}
The GR is governed by Einstein equation 
\begin{equation}
R_{\mu\nu}-\frac{1}{2}g_{\mu\nu}R=-\frac{8\pi{G}}{c^4}T_{\mu\nu}  \label{S1}
\end{equation}
corresponding to Einstein-Hilbert Lagrangian 
${\cal L}_{GR}=R$, where $R$ is the scalar curvature, $R_{\mu\nu}$ is the Ricci tensor and $T_{\mu\nu}$ 
is energy-momentum tensor. The gravitational wave propagation, in essence, distorts the space-time. 
Usually, this distortion is very weak, therefore, we can develop the general metric $g_{\mu\nu}$
peturbatively around the flat (Minkowski) metric $\eta_{\mu\nu}$, 
\begin{equation}
 g_{\mu\nu} = \eta_{\mu\nu} + h_{\mu\nu},    
 \label{metric}
\end{equation}
with $|h_{\mu\nu}|\ll 1$. 
Thus, $h_{\mu\nu}$ satisfies the equation
\begin{equation}
\square (h_{\mu\nu}-{1\over 2}\eta_{\mu\nu}\eta^{\alpha\beta}h_{\alpha\beta}) 
= 2kT_{\mu\nu},  ~~ k= \frac{8\pi{G}}{c^4}  
\label{grwe}
\end{equation}
which in vacuum becomes 
\begin{equation}
\square h_{\mu\nu} = 0.      
\end{equation}
It is the equation of gravitational waves propagating outside the source. 
They are linear combinations of two modes of (tensor) polarizations: the plus (+) 
and the cross ($\times$) ones\cite{Will:2018bme}. 
In the GR \cite{Weinberg1} the power of gravitational radiation is given by 
\begin{equation} 
P=-\frac{c^5}{8\pi G}\int_S \partial_j Q^{j 0i}dS^i= -\frac{c^5}{8\pi G}\int_V \partial_j \partial_i Q^{j 0i}dV, \label{S14}
\end{equation}
where  $ i, j=1, 2, 3 $ and $dS^i=n^i r^2 d\Omega$ with $d\Omega=\sin \theta d\theta d\varphi $,  
$ n^i=\displaystyle\frac{x^i}{r}$, while  
\begin{equation}
Q^{j 0i}=\frac{1}{2}\left(\delta^{ij}\partial^0 g^{kk}-\delta^{ij}\partial^k g^{0k}+\partial^j g^{i0}-\partial^0 g^{ij}\right). \label{S15}
\end{equation}
With the metric perturbation \eqref{metric}
the formula \eqref{S15}  becomes 
\begin{equation}
Q^{j 0i}=\frac{1}{2}\left(\delta^{ij}\partial^0 h^{kk}-\delta^{ij}\partial^k h^{0k}+\partial^j h^{i0}-\partial^0 h^{ij}\right). \label{Sh15}
\end{equation}
It will be shown below, in Sect. 4, that $Q^{j 0i}$ is related to the moment of inertia.
In a central field (no matter stationary or not) $P$ always vanishes, hence, no gravitational wave can be emitted.  The situation is different in the $f(R)$-gravitation because there the metric perturbation $h_{\mu\nu}$ is no longer stationary under the spherically symmetry condition as in the GR. 
Let us consider this in more details. 
\section{Gravitational radiation in the $f(R)$-theory of gravitation}
The $f(R)$-modified gravitation theory is based on the Lagrangian ${\cal L}= f(R)$, where $f(R)$ in general is an arbitrary (but well-defined) function of the scalar curvature $R$. The gravitational equation extending \eqref{S1} and corresponding to this Lagrangian has the form \cite{DeFelice:2010aj, Sotiriou:2008rp, Capozziello:2011et}  
\begin{align}
&f'(R)R_{\mu\nu}-g_{\mu\nu}\square f'(R)+\nabla_{\mu}\nabla_{\nu} f'(R)
-\frac{1}{2}f(R)g_{\mu\nu}\nonumber\\
&=-kT_{\mu\nu}.
\label{S16}
\end{align}
\noindent 
This theory is one of the simplest generations of the GR and takes the latter as a special case at $f(R)=R$  when  Eq.\eqref{S16} becomes Einstein equation \eqref{S1}. As is well known, to solve Einstein equation is always problematic because of its high non-linearity, but the equation \eqref{S16} is even more complicated. To simply the problem, the perturbation method for solving highly nonlinear equations is often used. Choosing this method is based on the following logical argument. Since Einstein's GR has already been tested as a very accurate theory (see in this context, for example, recent results \cite{MICROSCOPE:2022doy,Smetana:2022hid}), any theory extending the GR should deviate slightly from the latter, at least in the tested domains in the present epoch. That means ${\cal L} =f(R)$ should differ from Einstein-Hilbert  Lagrangian ${\cal L}_{GR}= R$ just slightly,  
\begin{equation}
f(R)=R+\lambda \underline{h}(R), \label{S17}
\end{equation}
where $\lambda$ and function $\underline{h}(R)$ must satisfy the condition 
$|\lambda \underline{h}(R)|\ll R$. \footnote{Note that here we change the notation $h(R)$  used in previous works \cite{KKV,VanKy:2022itq} to $\underline{h}(R)$ to avoid confusion with the trace of $h_{\mu\nu}$.} With this assumption, we can obtain a perturbative vacuum solution of Eq. \eqref{S16} for a central field  (in a Schwarzschild-type metric) {\color{blue} \cite{KKV}}
\begin{align}
ds^2= & \left[ 1-\frac{2GM_f(t)}{c^2r}\right] {dx^0}^2-\frac{dr^2}{1-\frac{2GM_f(t)}{c^2r}}
\nonumber \\ 
& -r^2(d\theta^2+\sin^2\theta{d\varphi^2}). \label{S18}
\end{align}
Note that the ``event horizon" happens at the modified Schwarzschild radius 
\begin{equation}
R_S = \frac{2GM_f}{c^2} 
\end{equation}
corresponding to something like an ``effective" mass,
\begin{align}
M_f(t):=M-\lambda M_1(t)-\lambda M_2(t),  \label{S19}
\end{align}
and thus an ``effective" black hole, with $G$ and $M$ being the gravitational constant and the (ordinary) mass of the 
gravitational source, respectively, while      
\begin{align}
&M_1(t)=\frac{2\pi [R_0(t)]^3}{3kc^2}\left[\underline{h}(kT^0_{~0})+kT^0_{~0}\underline{h}'(kT^0_{~0})\right],  \label{S20}
\\&
M_2(t)=\frac{4\pi}{kc^2}\underline{h}''(kT^0_{~0})\left[ \frac{\partial}{\partial t}\frac{M}{[R_0(t)]^3}\right]^2 ~\alpha (t),  \label{S21}
\\&
T^0_{~0}= {M c^2\over {\frac{4}{3}}\pi {[R_0(t)]}^3}, \label{S22}
\end{align}
where  $R_0(t)$ is the radius of the gravitational source in the moment $t$ (the subscript "$0$" prevents the radius $R_ 0$ from being confused with the Ricci scalar $R$), and  
\begin{align}
\alpha (t)=&\frac{3k^2c^2R_0(t)}{256\pi^2[\xi (t)]^4}\left\lbrace \frac{3}{\xi(t)R_0(t)}\arcsin[\xi (t) R_0(t)]
\right.
\nonumber
\\&
 -\left. \left(3+2[\xi(t)R_0(t)]^2\right) \sqrt{1-[\xi(t)R_0(t)]^2}\right\rbrace
\nonumber
\\&
\times \left( 1-[\xi (t)R_0(t)]^2\right)^{-3/2}, \label{S23}
\end{align}
with 
\begin{align}
\xi^2 (t)=\frac{2GM}{c^2 [R_0(t)]^3}. \label{S24}
\end{align}
As seen in \eqref{S18}, under the spherically symmetry condition, the metric perturbation $h_{\mu\nu}$, in particular, the components $h_+$ and $h_{\times}$,  generally do not vanish and are functions of the spacetime (in the GR with the spherical symmetry, they are stationary), where, $h_+ = (h_{11}-h_{22})/2$  and  $h_{\times}=h_{12}\equiv h_{21}$ are amplitudes of the plus- and the cross tensor polarizations, respectively.  Furthermore, the quantity \eqref{S15} , i.e., \eqref{Sh15}, and the radiation power \eqref{S14} in general do not vanish either. Apart from its magnitude satisfying the perturbation condition $|\lambda \underline{h}(R)|\ll R$, a general perturbation $\underline{h}(R)$ (of the theory itself, not the metric) does not generate $h_+ $ and $h_{\times}$ (or any element of $h_{\mu\nu}$) vanishing.  Consequently,  the general $f(R)$ theory, unlike the GR, allows gravitational waves of tensor modes, even under the spherically symmetry condition. Gravitational waves of scalar modes are also allowed,  but in some circumstance they are strongly suppressed \cite{Isi:2017fbj, LIGOScientific:2018czr,Isi:2022mbx, Inagaki:2023tjh}, 
therefore, in such a case they contribute little to the total radiation power. 
In the moment, we have not yet known the exact fraction of each mode's contribution to the total radiation.  It could be subject to further investigation.
%
\section{Testing models}
Let us see explicit expressions of $M_1(t)$ and $M_2(t)$ in some specific models considered in \cite{KKV,VanKy:2022itq}. In the model 
\begin{equation} 
f(R)=R+\lambda R^b; ~~ b>0,   \label{S25}
\end{equation}
$M_1(t)$ and $M_2(t)$ have the form 
\begin{align}
&M_1(t)=\frac{4\pi}{kc^2}\frac{(b+1)c^{2b}(kM)^b }{3^{1-b}~2^{2b+1}~{\pi}^b[R_0(t)]^{3b-3}}, 
\label{S26}\\
&M_2(t)=\frac{4\pi}{kc^2}\frac{b(b-1)c^{2b-4}(3kM)^{b-2}\left[ \frac{\partial}{\partial t}\frac{M}{[R_0(t)]^3}\right]^2 \alpha (t)}{(4\pi)^{b-2}[R_0(t)]^{3b-6}}, \label{S27}
\end{align}
while in the model 
\begin{equation}
f(R)=R^{1+\varepsilon} \label{S28}
\end{equation}
(with $|\varepsilon|\ll 1  $),  they are given as follows   
\begin{align}
&\lambda M_1(t)=-M+\frac{4\pi}{kc^2}\frac{ (\varepsilon +2)6^\varepsilon (kc^2M)^{\varepsilon +1}}{(8\pi)^{(\varepsilon +1)}[R_0(t)]^{3\varepsilon}},  \label{S29}\\
&\lambda M_2(t)=\frac{4\pi}{kc^2}\frac{\varepsilon (\varepsilon +1) (3kc^2M)^{\varepsilon -1}\left[ \frac{\partial}{\partial t}\frac{M}{[R_0(t)]^3}\right]^2 \alpha (t)}{(4\pi)^{\varepsilon -1}[R_0(t)]^{3\varepsilon -3}}. \label{S30}
\end{align}
The fact that $R_0(t)$ is a function of time while the star deforming (pulsating, expanding or contracting), leads to an  explicit time dependence of the metric (unlike the GR, where, a spherically symmetric solution is always stationary) and, thus, gravitational radiations are possible. We will see below how it happens.\\

Let us consider the $f(R)$-theory in \eqref{S17}, 
$$f(R)=R+\lambda \underline{h}(R),\eqno\eqref{S17}$$
then the Eq. \eqref{S16} becomes 
\begin{align}
R^{\mu}_{~\nu} &-\frac{1}{2}\delta^{\mu}_{~\nu}R+\lambda \underline{h}'(R)R^{\mu}_{~\nu}-\frac{\lambda}{2}\delta^{\mu}_{~\nu}
\underline{h}(R)-\lambda \delta^{\mu}_{~\nu}\square \underline{h}'(R) \nonumber \\ 
&+\lambda \nabla^{\mu}\nabla_{\nu}\underline{h}'(R)=-kT^{\mu}_{~\nu}. \label{pts1}
\end{align}
Inserting the solution of Einstein equation 
$$R=kT, ~~  R^{\mu}_{~\nu}=-k(T^{\mu}_{~\nu}-\frac{1}{2}\delta^{\mu}_{~\nu}T),$$
in perturbative terms of Eq. \eqref{pts1},  we get a perturbation equation of first order  
\begin{equation}
R^{\mu}_{~\nu}-\frac{1}{2}\delta^{\mu}_{~\nu}R=-k T^\mu_{f~\nu}, \label{pts2}
\end{equation}
with    
\begin{align}
	T^\mu_{f~\nu}= ~&T^{\mu}_{~\nu}-\lambda \underline{h}'(kT)(T^{\mu}_{~\nu}
	-\frac{1}{2}\delta^{\mu}_{~\nu}T) -\frac{\lambda}{2k}\delta^{\mu}_{~\nu}\underline{h}(kT) \nonumber\\
	&-\frac{\lambda}{k} \delta^{\mu}_{~\nu}\square^E \underline{h}'(kT)+\frac{\lambda}{k} \nabla^{\mu}\nabla_{\nu}^E \underline{h}'(kT)
 \nonumber\\
 = ~&  T^{\mu}_{~\nu} + \delta T^{\mu}_{~\nu} 
 \label{pts3}
\end{align}
treated as an "effective" (or modified) energy-momentum tensor, where,  
$\displaystyle \underline{h}'(kT)=\frac{\partial \underline{h}(kT)}{\partial (kT)}$ 
and the superscript $E$ in the covariant derivatives indicates that the metric tensor  $g_{\mu\nu}$ is taken in the Einstein equation solutions. \\


To solve \eqref{pts2} let us use the metric $g_{\mu\nu}$ split as in \eqref{metric} into flat and curved parts: 
%
$$
g_{\mu\nu}=\eta_{\mu\nu} + h_{\mu\nu},
\label{pts7}
\eqno\eqref{metric}
$$
The curved part $h_{\mu\nu}$ in turn is split into two parts,     %
\begin{equation}
h_{\mu\nu} = h^{(E)}_{\mu\nu} + h^{(f)}_{\mu\nu}, \label{pts9}
\end{equation}
of $h^{(E)}_{\mu\nu}$ - deviations within Einstein GR, and   
$h^{(f)}_{\mu\nu}$ - (perturbative) corrections by the $f(R)$-theory.
Next, following the approach of \cite{Landau1}, we obtain at the first order of perturbation
\begin{equation}
R_{\mu\nu}=\frac{1}{2}\left(g^{(0)\alpha\beta}\partial_\alpha\partial_\beta h_{\mu\nu}-\partial_\mu\partial_\sigma h^\sigma_\nu-\partial_\nu\partial_\sigma h^\sigma_\mu+\partial_\mu\partial_\nu h \right), \label{pts12}
\end{equation}
where $h=h^\mu_\mu$. 
Then, we can always choose a coordinate frame to 
satisfy 
\begin{equation}
\partial_\mu \chi^\mu_\nu=0, \label{pts13}
\end{equation}
where $\chi^\mu_\nu = h^\mu_\nu-\frac{1}{2}\delta^\mu_\nu h$. 
Using 
\eqref{pts13} 
we rewrite 
\eqref{pts12} as  
\begin{equation}
R_{\mu\nu}=\frac{1}{2}\square h_{\mu\nu}, \label{pts15}
\end{equation}
or
\begin{equation}
R^\mu_{~\nu}-\frac{1}{2}\delta^\mu_\nu R=\square \Upsilon^\mu_{~\nu}. \label{pts16}
\end{equation}
%
with $\square = \eta^{\mu\nu}\partial_\mu\partial_\nu
$, and 
\begin{equation}
\Upsilon^\mu_{~\nu}=\frac{1}{2}\left(h^\mu_{~\nu}-\frac{1}{2}\delta^\mu_\nu h \right). \label{pts17}
\end{equation}
Taking \eqref{pts2} into account we re-write Eq. \eqref{pts16} as a wave equation  
\begin{equation}
\square \Upsilon^\mu_{~\nu}=-k T^\mu_{f~\nu}. \label{pts18}
\end{equation}
with the source being the modified energy-momentum $T^\mu_{f~\nu}$. 
It is clear that at $\lambda h(R)=0$, the equation \eqref{pts18} reduces to that in \eqref{grwe}. 
The radiation power is calculated by formula \eqref{S38} or \eqref{S39} derived below.
With the standard procedure for solving  the wave function \eqref{pts18} we obtain the retarded solution 
\begin{equation}
\Upsilon^{\mu\nu}(\mathbf{x},t)= - {k\over 2\pi}\int {T_f^{\mu\nu}(\mathbf{x}', r-ct)\over |\mathbf{x}-\mathbf{x}'|}d^3\mathbf{x}',  \label{wsol}
\end{equation}
where we use the notations $x_\mu \equiv (x_0, x_i)= (ct,\mathbf{x})$ and $r=|\mathbf{x}|$ for the observer's coordinates and $x'_\mu$ for the coordinates of points within the source.  In the case of the GR, the solution  \eqref{wsol} becomes 
\begin{equation}
\Upsilon^{\mu\nu}(\mathbf{x},t)= - {k\over 2\pi}\int {T^{\mu\nu}(\mathbf{x}', r-ct)\over |\mathbf{x}-\mathbf{x}'|}d^3\mathbf{x}',  \label{wGW}
\end{equation}
with $T_f^{\mu\nu}$ replaced by $T^{\mu\nu}$. Far from the source 
(that is $|\mathbf{x}|\gg |\mathbf{x}'|$ or $|\mathbf{x}-\mathbf{x}'|\approx |\mathbf{x}| :=r$) and if the conservation law 
\begin{equation} 
\partial_\mu T^{\mu\nu}=0
\label{conslaw}
\end{equation}
is applied, we get (see \cite{Landau1}, and also \cite{Inagaki:2023tjh})    
\begin{align}
\Upsilon^{ij}(\mathbf{x},t) &= - {k\over 2\pi r}\partial_0\partial_0\int T^{00}(\mathbf{x}', r-ct)x'^ix'^jd^3\mathbf{x}', \nonumber \\
&= - {k\over 2\pi c^2r} {d^2 {\cal I}^{ij}\over d^2t},
\label{wGW2}
\end{align}
where ${\cal I}^{ij}$ is the moment of inertia. 
The radiation power is given by \cite{Landau1} 
\begin{equation}
 P= - {G\over 45c^5}\left({\partial^3\over \partial t^3}{\cal Q}_{ij}\right)^2, 
 \label{epow}
\end{equation}
where 
\begin{equation}
 {\cal Q}_{ij}= \int_V \rho(\mathbf{x}')(3x'_ix'_j-|\mathbf{x}'|^2\delta_{ij})d^3x'.
 \label{4-q}
\end{equation}
Here ${\cal Q}_{ij}$ is the quadrupole moment and the integration is done over the volume (denoted by $V$) of the source.  It is easy to show that with the  spherical symmetry present, ${\cal Q}_{ij}$ vanishes, hence, there is no gravitational radiation for a central field. In the case of the $f(R)$-theory, however, the situation is different.\\

\begin{figure}[h]
    \centering
    \includegraphics[scale=1.8]{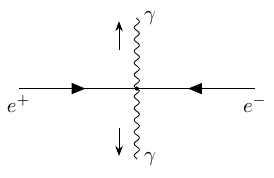}
    \caption{\textit{Two-gamma production of an electron-positron collision}.}
    \label{2gamma}
\end{figure}

Indeed, in the latter theory, we cannot impose the equation \eqref{conslaw} on $T_f^{\mu\nu}$, as, in general, 
\begin{equation}
\partial_\mu T_f^{\mu\nu}=\partial_\mu T^{\mu\nu} + \partial_\mu(\delta T^{\mu\nu})= \partial_\mu(\delta T^{\mu\nu})\neq 0,    
\end{equation}
that is, the effective energy-momentum tensor $T_f^{\mu\nu}$ (in particular, the effective mass $M_f$), unlike the real energy-momentum tensor $T^{\mu\nu}$ (in particular, the real mass $M$, resp.), is not necessarily conserved. It is so because the formulas \eqref{epow} and \eqref{4-q}  get corrections beyond the quadrupole terms, hence, neither the corrected ${\cal Q}^f_{ij}$  nor the corrected $P^f$ vanish for a central field. These results, along with the non-vanishing $h_+$ and $h_{\times}$ derived from the spherically symmetric time-varying metric \eqref{S18}, show that gravitational radiation of tensor mode is possible in the condition of spherical symmetry. This is the subject of the wave equation \eqref{pts18} solved by the solution \eqref{wsol}. What about the angular momentum conservation for the tensor (spin 2) mode radiation, we always have it because the radiation is spherically symmetric, i.e., same in all directions, in particular, in two opposite directions. This situation is similar to, for example,  the two-photon production in a head-on collision between a particle and its antiparticle, such as a head-on electron-positron collision, observed in a center-of-mass frame of reference, as depicted in Fig.  \ref{2gamma}.
\\

As 
$dr=\displaystyle\frac{\overrightarrow{r}d\overrightarrow{r}}{r} =-\frac{x_idx^i}{r}$, thus,  
$dr^2=\displaystyle\frac{x_ix_jdx^idx^j}{r^2}$, 
at a location far from the gravitational 
source the metric \eqref{S18} 
\noindent
in a Cartesian coordinate frame takes the form
\begin{equation}
ds^2=\left[ 1-\frac{2GM_f(t)}{c^2r}\right] {dx^0}^2-\left[ \delta_{ij}+\frac{2GM_f(t)}{c^2}\frac{x_ix_j}{r^3}\right] dx^idx^j. \label{S34}
\end{equation}
In this perturbative $f(R)$-theory the formula for radiation has a similar form \eqref{S14}--\eqref{S15} as in the GR with the difference that the metric $g_{\mu\nu}$ is stationary in the GR but time-dependent in the $f(R)$-theory.  
Thus, the tensor polarization is reflected by 
$$h_+ = \frac{GM_f(t)}{c^2r^3}[(x_1)^2-(x_2)^2]$$ and 
$$h_\times =\frac{2GM_f(t)}{c^2r^3}x_1x_2,$$
while $Q^{j0i}$ in \eqref{S15} becomes     

\begin{align}
Q^{j0i} &=-{G\dot{M}_f(t)\over c^3}\left({\delta^{ij}\over r}-{x^ix^j\over r^3}\right)
\nonumber \\ 
& =-{4\pi G\dot{\rho}_f(t)\over 3c^3}\left(r^2\delta^{ij}-x^ix^j\right),  \label{S35}
\end{align}
where, 
\begin{equation}
\dot{M}_f(t)\equiv \frac{\partial M_f(t)}{\partial t} =\dot{M}(t)-\lambda \dot{M}_1(t)-\lambda \dot{M}_2(t),   \label{S36}
\end{equation} 
according to \eqref{S19} and 

\begin{equation}
\rho_f(t)=\frac{3M_f(t)}{4\pi r^3}
\label{rho_r}
\end{equation}
is the effective density of the matter inside the sphere with radius $r$.  
We see from \eqref{S35} that $Q^{j0i}$ is proportional to the time derivative of the (effective) density of the moment of inertia. If we calculate the density only within the source with radius $R_0$ the formula \eqref{S35} 
is replaced by 
\begin{equation}
Q^{j0i}= - \frac{4\pi G\dot{\rho}^{(s)}_f (t) R^3_0}{3c^3r^3}\left(r^2\delta^{ij}- x^ix^j\right) , \label{S35b}
\end{equation}
where 
\begin{equation}
\rho^{(s)}_f (t)=\frac{3M_f(t)}{4\pi R_0^3}
\label{rho_r}
\end{equation}
is the effective density of the source. \\

Using $r=\sqrt{-x_ix^i}$,  $\partial_ir=-x_i/r$ we get
\begin{equation}
\partial_jQ^{j0i}=\frac{2G\dot{M}_f(t)}{c^3}\frac{x^i}{r^3}. \label{S37}
\end{equation}
Inserting \eqref{S37} in \eqref{S14}, we obtain the gravitation radiation power of a central field in the $f(R)$-gravitation  
\begin{equation}
P_f=-\dot{M}_f(t)c^2. \label{S38}
\end{equation}
With a spherical symmetry we can consider $\dot{M}(t)\approx 0$ since the total real mass $M$, compared with $M_1$ and $M_2$, is a constant or slowly-varying function of time (thus, any possible effect of the pure GR is excluded). Hence, the main contribution to \eqref{S36} comes from the last two terms, that is   
\begin{equation}
P_f=\lambda \left[ \dot{M}_1(t)+\dot{M}_2(t)\right] c^2. \label{S39}
\end{equation}
 To verify \eqref{S39} experimentally, let us make some  numerical estimations of gravitational radiation based on \eqref{S39} in several cases of gravitational collapse. To increase the observable (measurable) effect, along with a large mass $M$ a rapid change with time of $M_f(t)$, thus, 
after \eqref{S20} and \eqref{S21}, 
of $R_0(t)$, is required. Therefore, within the $f(R)$-theory, a rapid change of $R_0(t)$ may cause a detectable gravitational radiation. \\

 When a (heavy) object deforms very quickly, such as in the case of a gravitational collapse, the contribution from $\dot{M}_2(t)$ would dominate  the other contributions. In such a case the radiation formula \eqref{S39} reduces to 
 \begin{equation} 
P=\lambda\dot{M}_2(t)c^2. \label{S39a}
\end{equation} 
Then, the total radiation energy released during a time interval 
$\bigtriangleup t=t-0=t$, 
\begin{equation} 
E_{rad}=\int_0^t Pdt=\lambda c^2\int_0^t dM_2(t)
 = \lambda c^2\bigtriangleup(M_2(t)),  \label{S39b}
\end{equation} 
is measured by the change  $\bigtriangleup(M_2(t))=M_2(t)-M_2(0)$ of the ``fictitious" mass $M_2$. This is a way geometry is converted into mass-energy. Whether energy of this kind can be treated as dark energy is still a matter for further consideration. The difference with the GR is that, in the latter, unlike \eqref{S39b}, the radiation energy is measured by the reduction ("evaporation") of the real total mass $M$ (of, e.g., colliding objects).\\
 
\begin{figure}[h]
\centering	
	\includegraphics[scale=0.35]{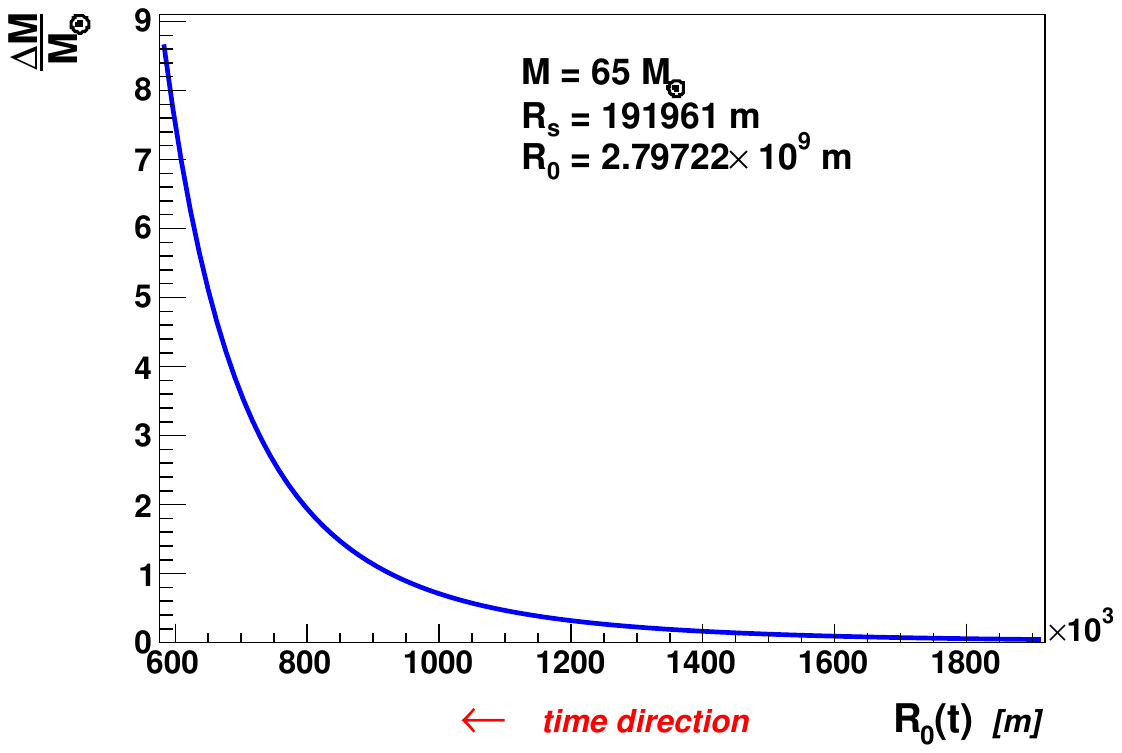}
	\caption{\label{M65}\textit{The change of the "fictitious" mass due to gravitational radiation of a 65-solar-mass object (e.g., a star) collapsing from a size with a radius of 10 its Schwarzschild radius $R_s$ to 3 $R_s$ .}}
\end{figure}

\begin{figure}[h]
	\centering	
	\includegraphics[scale=0.35]{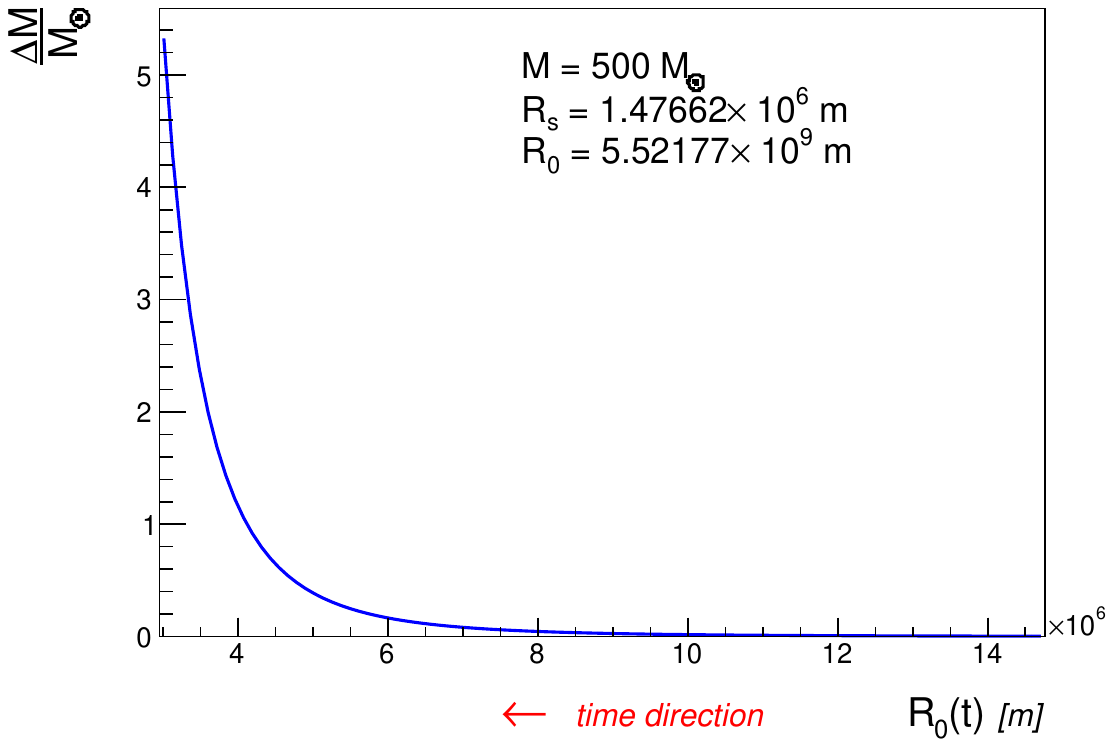}
	\caption{\label{M500}\textit{The change of the "fictitious" mass due to gravitational radiation of a 500-solar-mass object collapsing from a size with a radius of 10 its Schwarzschild radius $R_s$ to 2 $R_s$.}}
	\end{figure}

\begin{figure}[h]
	\centering		
	\includegraphics[scale=0.35]{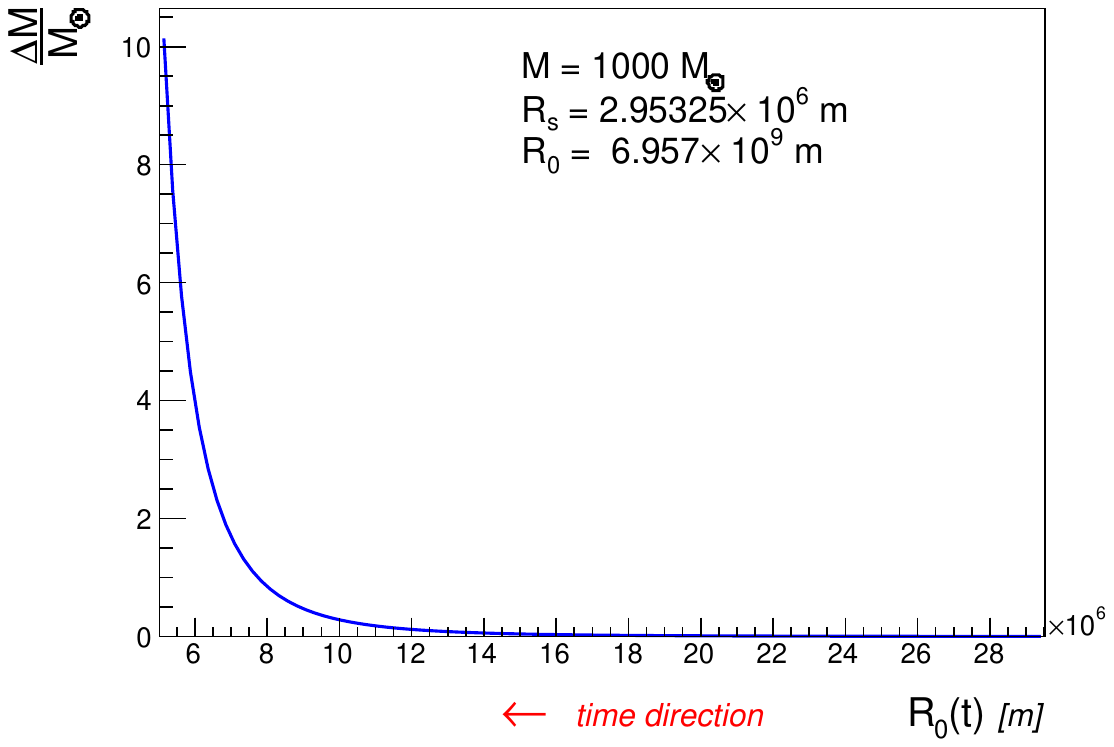}
	\caption{\label{M1000}\textit{The change of the "fictitious" mass due to gravitational radiation of a 1000-solar-mass object collapsing from a size with a radius of 10 its Schwarzschild radius $R_s$ to 1.7 $R_s$.}}
\end{figure}

\begin{figure}[h]
 \includegraphics[scale=0.35]{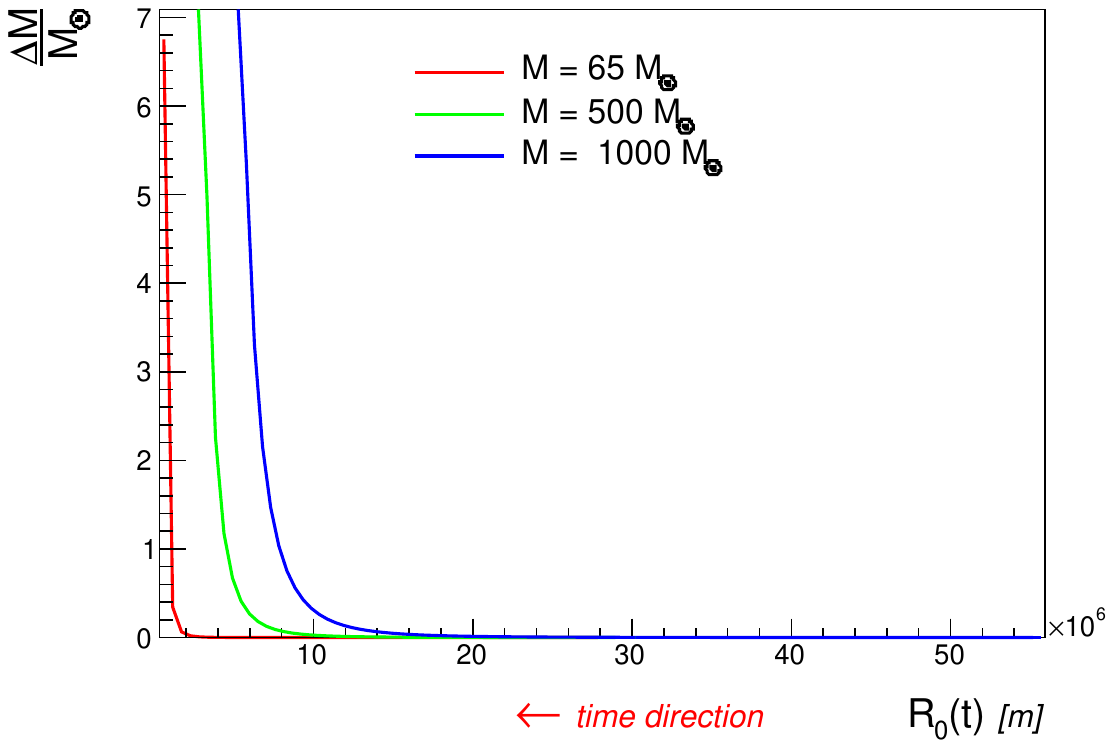}
	\caption{\label{Complot}\textit{Combined plot of three cases $M=65,500, 1000 M_\odot$. In all cases, the gravitational radiation flares up as the star's surface approaches, without reaching, the (modified) event horizon.}}
\end{figure}

Inserting the latter formula in \eqref{S21} we can obtain $M_2$, and then, the radiation power and the total radiation energy by \eqref{S39a} and \eqref{S39b}, respectively, for a general $f(R)$. For a specific model, a formula of a specific type such as \eqref{S27} or \eqref{S30} should be taken instead of the general one \eqref{S21}. To make the imagination more intuitive, now we apply this procedure to a model with a definite $f(R)$, namely, the model \eqref{S25} with $b=2$, that is, 
\begin{equation}
f(R)=R+ \lambda R^2,
\label{fR_test}
\end{equation}
with $\lambda = 0.1511677 \times 10^{18}~m^2$ given in \cite{KKV}.
 Denoting $\Delta M_f =\displaystyle\frac{E_{rad}}{c^2}$ 
(but below, for brevity, the script $f$ will be omitted: 
$\Delta M_f \equiv \Delta M$)
and using \eqref{S39b} and \eqref{S27} (with $b=2$)
we find 
\begin{align}
\frac{\Delta M}{M_\odot}=&\frac{27\lambda c^2}{32G^2 MM_\odot}\left( \frac{\partial R_0(t)}{\partial t}\right)^2
\nonumber
\\&
\times
\left[ \frac{3}{\sqrt{\frac{2GM}{c^2R_0(t)}}}\arcsin\sqrt{\frac{2GM}{c^2R_0(t)}} 
\right.
\nonumber
\\
&
\left.
-2\left( \frac{3}{2}+\frac{2GM}{c^2R_0(t)}\right)\sqrt{1-\frac{2GM}{c^2R_0(t)}} \right]\frac{2GM}{c^2R_0(t)}
\nonumber
\\
&\times
\left( 1-\frac{2GM}{c^2R_0(t)}\right)^{\frac{-3}{2}}. \label{c1a}
\end{align}
To illustrate the above-derived result, let us consider a collapsing star. We assume the star's gravitational collapse considered as a free fall of its constituent matter toward its center. Likely, this can happen during a late stage of the star's evolution when the star becomes very cool (so its fuel is exhausted to counteract a gravitational collapse). Thus, $dt$ can be calculated as follows \cite{star-evol}  
\begin{eqnarray} 
dt=-\left(\frac{8\pi G\rho_0}{3} \right)^{-1/2}\left(\frac{\zeta}{1-\zeta} \right)^{1/2}d\zeta,\label{dt1} 
\end{eqnarray}
where $ \zeta=\displaystyle\frac{R_0(t)}{R_0} $, $ \rho_0=\displaystyle\frac{M}{\displaystyle 4/3\pi \left( R_0\right) ^3} $,  $R_0$ is the radius of the collapsing star at a given time just before the collapse, taken as the initial time ($t=0$), while $R_0(t)$ is the radius of the star at a later time $t$. 
%
%
Following \eqref{dt1}, the speed of the collapse can be approximated as 
\begin{equation}
\frac{\partial R_0(t)}{\partial t}=-\sqrt{\frac{2GM}{R_0(t)R_0}[R_0-R_0(t)]}. \label{dt2}
\end{equation}
Putting (\ref{dt2}) in (\ref{c1a}), we finally find  
\begin{align}
\frac{\Delta M}{M_\odot}=&\frac{27\lambda c^2\left[ R_0-R_0(t)\right] }{16GM_\odot R_0(t)R_0}\left[ \frac{3}{\sqrt{\frac{2GM}{c^2R_0(t)}}}\arcsin\sqrt{\frac{2GM}{c^2R_0(t)}}
\right.
\nonumber
\\
&\left.
-2\left( \frac{3}{2}+\frac{2GM}{c^2R_0(t)}\right)\sqrt{1-\frac{2GM}{c^2R_0(t)}} \right]
\nonumber
\\
&\times
\frac{2GM}{c^2R_0(t)}\left( 1-\frac{2GM}{c^2R_0(t)}\right)^{\frac{-3}{2}}. \label{dt3}
\end{align}
We work here on three testing examples with $M=65 M_\odot$, $M=500 M_\odot$ and  $M=1000 M_\odot$. 
The radius $R_0$ of a star of mass $M$ assumed to have a prio-collapse density like that of the Sun, is, thus,  approximately given by $R_0 = \left( M/M_\odot\right)^{1/3} R_{\odot}$.  If  $R_0(t)\ll R_0$ (i.e., when $t$ is large enough), the precise choice of $R_0$   does not affect expression \eqref{dt3} much. 
The case with $M=65 M_\odot$ is considered in order to collate it with the LIGO's observation \cite{LIGOScientific:2016aoc}. As said above, the gravitational waves in an $f(R)$-theory can be released by virtue of increasing "fictitious" $M_2$ mass, unlike in the conventional GR the gravitational radiation leads to reduction of the conventional (real) mass $M$. This phenomenon is depicted in Figs. \ref{M65}, \ref{M500} and \ref{M1000} for  $M=65 M_\odot$, $M=500 M_\odot$ and $M=1000 M_\odot$, respectively. 
In these figures where $\Delta M \equiv \Delta (M_2)$, we see a (positive) change of $M_2$ happens due to a gravitational radiation, or, a change of the mass $M_2$ causes a gravitational radiation. The Fig. \ref{Complot} depicts a combined plot to compare the three above-considered cases. It is observed that when the radius of the collapsing star approaches its Schwarzschild-like radius $R_S$ the gravitational radiation flares up. 
The model \eqref{fR_test} is just a testing model which can be replaced by other ones, such as that with $f(R)=R-2\Lambda + \alpha R^2 +{\gamma\over R},$ shown in \cite{VanKy:2022itq} as a more viable model. We note that as  the Ricci tensor $R^\mu_{~\nu}$ (also the Riemann tensor), see \eqref{pts12}, and thus, the Ricci scalar $R$, even under the spherically symmetry condition, are not stationary (but functions of the space-time)  the tensor modes of gravitational waves are not cancelled out \cite{Eardley:1973br, Eardley:1973zuo}.
%
%
\section{Conclusion}
To conclude, let us make some remarks. It has been shown that in the $f(R)$-theory of gravitation the so-called Birkhoff theorem (valid in the GR) is overcome, or more precisely, in the spherically symmetric condition, this theory allows gravitational radiation of tensor modes and scalar modes, unlike the GR, which in general allows only radiation of tensor modes and does not even allow under the spherically symmetric condition any radiation of any mode.\\

In the GR, any  central gravitational field is always stationary, therefore, no gravitational radiation can be emitted from a spherically symmetric gravitational source.  In contrast, a spherically symmetric varying gravitational source in the $f(R)$ theory could emit gravitational waves detectable (on the Earth) if the variation is strong and quick enough. That means that gravitational waves can be radiated by a single varying spherical gravitational source (for example, a star in collapse/explosion or pulsation - not necessary to have a collision of two or more heavy objects such as black holes) - a phenomenon impossible in the GR. In other words, the $f(R)$-theory of gravitation predicts more possible sources of gravitational waves than the GR. This increases the chances of detection of  gravitational waves.  These waves can be of tensor or scalar mode, whereas the latter could be strongly suppressed, however, we do not yet know the exact fraction of each mode in the total radiation. We have calculated radiation power formulas that can be used to instruct a gravitational-wave-detecting experiment to verify the $f(R)$  theory of gravitation.  Measurements by such a potential experiment can fix the precise form of $f(R)$. 
The technical challenges, such as precision, sensitivity, etc. of the instruments as well as signal analysis to distinguish the signal with background (radiations from other sources), can be the biggest obstacle \cite{Smetana:2022hid,Jana:2022shb}. In particular, a technique \cite {Jana:2022shb} applied to detecting binary GR black holes, could be used for detecting single black holes in the $f(R)$-gravitation.\\

Finally, let us make some extended discussion. 
We note that for some models (e.g., those in \eqref{S25} with $0<b<1$) during a gravitational collapse the mass $M_1$ decreases, while the mass $M_2$ increases. A gravitational radiation occurs only if the change of $M_2$ surpasses the change of $M_1$. It is possible as a gravitational collapse usually happens very quickly. On the other hand, if an explosion of a star leads to an increasing $M_ 1$ and a decreasing $M_ 2$ (when $R_0(t)$ does not change fast enough) no gravitational radiation occurs, unless $\lambda<0$. For the case $\lambda<0$ (or $\lambda>0$ with fast changing $R_0(t)$) an explosion of a heavy object could lead to gravitational radiation. The plots for such processes are similar to those in Figs. \ref{M65}--\ref{Complot} but with the time direction reversed. In any deformation direction, the gravitational radiation is most intense just outside the event horizon. 
We speculate that if the deforming object is a larger one such as a galaxy or even the Universe itself we could consider the question of the origin of dark matter and dark energy. That means, a deformation of a cosmic object could ``produce" (or ``absorb")  energy treated as a contribution to the energy-matter budget (including dark energy and dark matter) of the Universe. It is geometry converted into energy-matter, according to some speculation \cite{Capozziello:2008rq}
\\[4mm]
\noindent
{\bf Acknowledgements:} 
The authors would like to thank the referee and
the editor-in-chief for valuable comments.
All the authors have contributed equally to this work. 
The work of N. T. H. V. is funded by Vietnam 
Academy of Science and Technology under Grant No. NVCC05.06/24-25.
\\

\noindent
{\bf Data Availability Statement:} 
This manuscript has no associated data or the data will not be deposited. 
[Authors' comment: This is a theoretical work and no experimental data].

\begin{thebibliography}{9}
%
\bibitem{Birkhoff1} G. D. Birkhoff and R. E. Langer, ``\textit{Relativity and modern
		physics}'', Harvard U. Press, Cambridge, MA, 1923.

\bibitem{VojeJohansen:2005nd} 
	N.~Voje Johansen and F.~Ravndal,
	``\textit{On the discovery of Birkhoff's theorem}'',
	Gen.\ Rel.\ Grav.\  {\bf 38}, 537 (2006);
	[physics/0508163].
	
\bibitem{Weinberg1} S. Weinberg, \textit{``Gravitation and cosmology: Principle and applications of the general relativity}'', John Wiley  $ \& $ Son, New York, 1972.
%

\bibitem{Landau1}L. D. Landau and  E. M. Lifshitz, ``\textit{The classical theory of fields}",  vol. 2, Elsevier, Oxford, 1994.
%

\bibitem{KKV}
N.~Anh Ky, P.~V.~Ky and N.~T.~H.~Van,
\textit{``Perturbative solutions of the $f(R)$-theory of gravity in a central gravitational field and some applications''},
Eur. Phys. J. C \textbf{78}, no.7, 539 (2018)
[erratum: Eur. Phys. J. C \textbf{78}, 664 (2018)];
[arXiv:1807.04628 [gr-qc]].
%

\bibitem{ReddyEPJP2018}
P.J. Ravindranath1, Y. Aditya2, D.R.K. Reddy3,a, and M.V. Subba Rao, 
``\textit{Birkhoff’s theorem in f(R) theory of gravity}'', 
Eur. Phys. J. Plus (2018) 133: 376. 

\bibitem{DeFelice:2010aj} 
  A.~De Felice and S.~Tsujikawa,
  ``\textit{f(R) theories}'',
  Living Rev.\ Rel.\  {\bf 13}, 3 (2010);
  [arXiv:1002.4928 [gr-qc]].

\bibitem{Sotiriou:2008rp}
T.~P.~Sotiriou and V.~Faraoni,
``\textit{f(R) Theories Of Gravity}'',
Rev. Mod. Phys. \textbf{82}, 451-497 (2010);
[arXiv:0805.1726 [gr-qc]].

\bibitem{Capozziello:2011et} 
  S.~Capozziello and M.~De Laurentis,
  ``\textit{Extended theories of gravity}'',
  Phys.\ Rept.\  {\bf 509}, 167 (2011);
  [arXiv:1108.6266 [gr-qc]].

\bibitem{LIGOScientific:2016aoc}
B.~P.~Abbott \textit{et al.} [LIGO Scientific and Virgo],
``\textit{Observation of gravitational waves from a binary black hole merger}'',
Phys. Rev. Lett. \textbf{116}, 061102 (2016);
[arXiv:1602.03837 [gr-qc]].
%

\bibitem{VanKy:2022itq}
P.~Van Ky, N.~T.~Hong.~Van and N.~Anh Ky,
``\textit{Perturbative approach to f(R)-gravitation in FLRW cosmology}'', 
Eur. Phys. J. C \textbf{83}, 330 (2023); 
[arXiv:2206.11259 [gr-qc]].
%

\bibitem{Katsuragawa:2019uto}
T.~Katsuragawa, T.~Nakamura, T.~Ikeda and S.~Capozziello,
``\textit{Gravitational waves in $F(R)$ gravity: scalar waves and the chameleon mechanism}'',
Phys. Rev. D \textbf{99}, 124050 (2019); 
[arXiv:1902.02494 [gr-qc]].
%

\bibitem{Gogoi:2020ypn}
D.~J.~Gogoi and U.~Dev Goswami,
``\textit{A new $f(R)$ gravity model and properties of gravitational waves in it}'', 
Eur. Phys. J. C \textbf{80}, 1101 (2020);
[arXiv:2006.04011 [gr-qc]].
%
\bibitem{Eardley:1973br}
D.~M.~Eardley, D.~L.~Lee, A.~P.~Lightman, R.~V.~Wagoner and C.~M.~Will,
``\textit{Gravitational-wave observations as a tool for testing relativistic gravity}", 
Phys. Rev. Lett. \textbf{30}, 884-886 (1973).
%
\bibitem{Eardley:1973zuo}
D.~M.~Eardley, D.~L.~Lee and A.~P.~Lightman,
``\textit{Gravitational-wave observations as a tool for testing relativistic gravity}'', 
Phys. Rev. D \textbf{8}, 3308-3321 (1973).
%
\bibitem{Capozziello:2008rq}
S.~Capozziello, C.~Corda and M.~F.~De Laurentis,
``\textit{Massive gravitational waves from f(R) theories of gravity: Potential detection with LISA}'', 
Phys. Lett. B \textbf{669}, 255-259 (2008);
[arXiv:0812.2272 [astro-ph]].
%
\bibitem{Nishizawa:2009bf}
A.~Nishizawa, A.~Taruya, K.~Hayama, S.~Kawamura and M.~a.~Sakagami,
``\textit{Probing non-tensorial polarizations of stochastic gravitational-wave backgrounds with ground-based laser interferometers}'', 
Phys. Rev. D \textbf{79}, 082002 (2009); 
[arXiv:0903.0528 [astro-ph.CO]].
%

\bibitem{Inagaki:2023tjh}
T.~Inagaki and M.~Taniguchi,
``\textit{Scalar mode quadrupole radiation from astronomical sources in F(R) modified gravity}'', 
Phys. Rev. D \textbf{108}, 024003 (2023);
[arXiv:2302.02734 [gr-qc]].
%

\bibitem{Isi:2017fbj}
M.~Isi and A.~J.~Weinstein,
``\textit{Probing gravitational wave polarizations with signals from compact binary coalescences}'',  
[arXiv:1710.03794 [gr-qc]].
%

\bibitem{LIGOScientific:2018czr}
B.~P.~Abbott \textit{et al.} [LIGO Scientific and Virgo],
``\textit{Search for tensor, vector, and scalar polarizations in the stochastic gravitational-Wave background}'', 
Phys. Rev. Lett. \textbf{120}, 201102 (2018); 
[arXiv:1802.10194 [gr-qc]].
%
\bibitem{Isi:2022mbx}
M.~Isi,
``\textit{Parametrizing gravitational-wave polarizations}'',
Class. Quant. Grav. \textbf{40}, 203001 (2023);
[arXiv:2208.03372 [gr-qc]].
%
\bibitem{Inagaki:2023tjh}
T.~Inagaki and M.~Taniguchi,
``\textit{Scalar mode quadrupole radiation from astronomical sources in F(R) modified gravity}'', 
Phys. Rev. D \textbf{108}, 024003 (2023); 
[arXiv:2302.02734 [gr-qc]].
%
\bibitem{Will:2018bme} C.~M.~Will, 
``\textit{Theory and experiment in gravitational physics}'',  Cambridge university press, 2018.
%

 \bibitem{MICROSCOPE:2022doy}
 P.~Touboul \textit{et al.} [MICROSCOPE],
 ``\textit{MICROSCOPE mission: Final results of the test of the equivalence principle}'',
 Phys. Rev. Lett. \textbf{129}, 121102 (2022).
%

\bibitem{Smetana:2022hid}
J.~Smetana, R.~Walters, S.~Bauchinger, A.~S.~Ubhi, S.~Cooper, D.~Hoyland, R.~Abbott, C.~Baune, P.~Fritchel and O.~Gerberding, \textit{et al.}
``\textit{Compact Michelson interferometers with subpicometer sensitivity}'',
Phys. Rev. Applied \textbf{18}, 034040 (2022);  
[arXiv:2202.10274 [physics.ins-det]].
%
\bibitem{star-evol} Kippenhahn R and Weigert A and Weiss A,  
``\textit{Stellar structure and evolution}'', Springer-Verlag, 2012.
%
\bibitem{Jana:2022shb}
S.~Jana, S.~J.~Kapadia, T.~Venumadhav and P.~Ajith,
``\textit{Cosmography using strongly lensed gravitational waves from binary black holes}'',
Phys. Rev. Lett. \textbf{130}, no.26, 261401 (2023); 
[arXiv:2211.12212 [astro-ph.CO]].

\end{thebibliography}
\end{document}